\documentclass[12pt]{article}

\usepackage{arxiv}

\usepackage{graphicx}
\usepackage{dcolumn}
\usepackage{bm}
\usepackage[mathlines]{lineno}
\usepackage{color}
\usepackage{todonotes}
\usepackage{amssymb}
\usepackage{amsmath}
\usepackage{upgreek}

\usepackage{authblk}

\usepackage[symbol]{footmisc}

\renewcommand{\headeright}{}
\renewcommand{\undertitle}{Preprint}

\title{Non-destructive characterisation of dopant spatial distribution in cuprate superconductors}

\author[1,2*]{A.-E. \c{T}u\c{t}ueanu}
\author[3*]{M. Sales }
\author[4,2]{K. L. Eliasen}
\author[5,2]{M.-E. L\v{a}c\v{a}tu\c{s}u}
\author[5]{J.-C. Grivel}
\author[6]{N. Kardjilov}
\author[6]{I. Manke}
\author[6]{M. Krzyzagorski}
\author[7]{Y. Sassa}
\author[9,8]{M. S. Andersson}
\author[3]{S. Schmidt}
\author[2]{K. Lefmann}

\affil[1]{Institute Max von Laue Paul Langevin, 38042 Grenoble, France}
\affil[2]{Nanoscience Center, Niels Bohr Institute, University of Copenhagen, 2100 Copenhagen {\O}, Denmark}
\affil[3]{Department of Physics, Technical University of Denmark, 2800 Kgs.\ Lyngby, Denmark}
\affil[4]{Glass and Time, IMFUFA, Department of Science and Environment, Roskilde University, DK-4000 Roskilde, Denmark}
\affil[5]{Department of Energy Conversion and Storage, Technical University of Denmark, 2800 Kgs.\ Lyngby, Denmark}
\affil[6]{Institute of Applied Materials, Helmholtz-Zentrum Berlin, 14109 Berlin, Germany}
\affil[7]{Department of Physics, Chalmers University of Technology, SE-412 96 G{\"o}teborg, Sweden}
\affil[8]{Department of Chemistry and Chemical Engineering, Chalmers University of Technology, 412 96 G{\"o}teborg, Sweden}
\affil[9]{Department of Engineering Sciences, Uppsala University, 751 21 Uppsala, Sweden}
\affil[*]{Corresponding authors: A.-E. \c{T}u\c{t}ueanu, anaelena@nbi.ku.dk and M. Sales, msales@fysik.dtu.dk}
\date{} 	

\begin{document}
\maketitle

\begin{abstract}
Proper characterisation of investigated samples is vital when studying superconductivity as impurities and doping inhomogeneities can affect the physical properties of the measured system. We present a method where a polarised neutron imaging setup utilises the precession of spin-polarised neutrons in the presence of a trapped field in the superconducting sample to spatially map out the critical temperature for the phase transition between superconducting and non-superconducting states. We demonstrate this method on a superconducting crystal of the prototypical high-temperature superconductor (La,Sr)$_2$CuO$_4$. The results, which are backed up by complementary magnetic susceptibility measurements, show that the method is able to resolve minor variations in the transition temperature across the length of the LSCO crystal, caused by inhomogeneities in strontium doping.
\end{abstract}

\section{Introduction}
More than 30 years after Bednorz and M\"uller's discovery of superconductivity (SC) at temperatures above 30 K \cite{bednorz1986possible}, the nature of these unconventional or high-temperature superconductors (HTSC) is still to be understood. Unlike conventional superconductors, where the electron pairing mechanism is well described through the theory put forward by Bardeen, Cooper and Schrieffer (BCS) \cite{schrieffer2018theory}, the pairing mechanism in the unconventional HTSC remains an unresolved question to this day. In recent years, increasing amounts of evidence points to the interplay between magnetism and superconductivity as a key factor in the emergence of the latter \cite{keimer2015quantum}. One of the most widely used methods of studying highly correlated electron systems, and in particular intertwined electronic orders, is neutron scattering. Especially when investigating spin density wave order and the related fluctuations in cuprate superconductors, neutron scattering has played a major part in revealing the nature of their relationship with SC \cite{tranquada2007neutron, tranquada2013spins, jacobsen2018distinct}.

One of the many advantages of using neutrons as a probe is their charge neutrality, which makes them interact rather weakly with most matter \cite{Squires}, in contrast with, for example, X-rays. This allows the probing of the entire bulk of the sample in different environments such as magnets, pressure cells, cryostats or furnaces. However, the drawback is that, in order to measure weak signals, samples of large masses are needed during experiments. When it comes to cuprate superconductors, the growth of large single crystals is not a trivial task. On top of that, due to the fact that the intrinsic properties of these materials are under scrutiny, the use of high quality and purity samples is a prerequisite as the analysed physical properties can be drastically affected by defects, impurities or inhomogeneities. This has increased the need for non-destructive characterisation methods which are able to accurately determine the physical properties of the entire bulk of the sample. One such property is the doping concentration, a parameter extremely important in defining the phase diagram of the compound, since it has direct impact on the superconducting properties \cite{kofu2009hidden}. 

In this paper, we demonstrate the use of polarised neutron imaging \cite{strobl2018topical, dawson2009imaging, kardjilov2018advances} as a non-destructive method for determining the distribution of the superconducting critical temperature (T$_\text{C}$) along the length of a HTSC sample, which is in turn related to the doping concentration  distribution. Our sample is the prototypical cuprate HTSC La$_{2-x}$Sr$_x$CuO$_4$ (LSCO). With this method we were able to detect an inhomogeneous distribution of the dopant throughout the sample. In contrast, when performing standard magnetic susceptibility measurements only small sample pieces are used, usually cut from only one side of the crystal, thus the result cannot be considered representative for the entire bulk. By measuring magnetic susceptibility on pieces originating from both ends of the main single crystal we were able to confirm the T$_\text{C}$ difference, but the precise Sr concentration at each position in the sample went undetected.


\section{The superconducting sample and its inhomogeneity level}
\begin{figure}[h!]
    \centering
    \includegraphics[width=0.7\columnwidth]{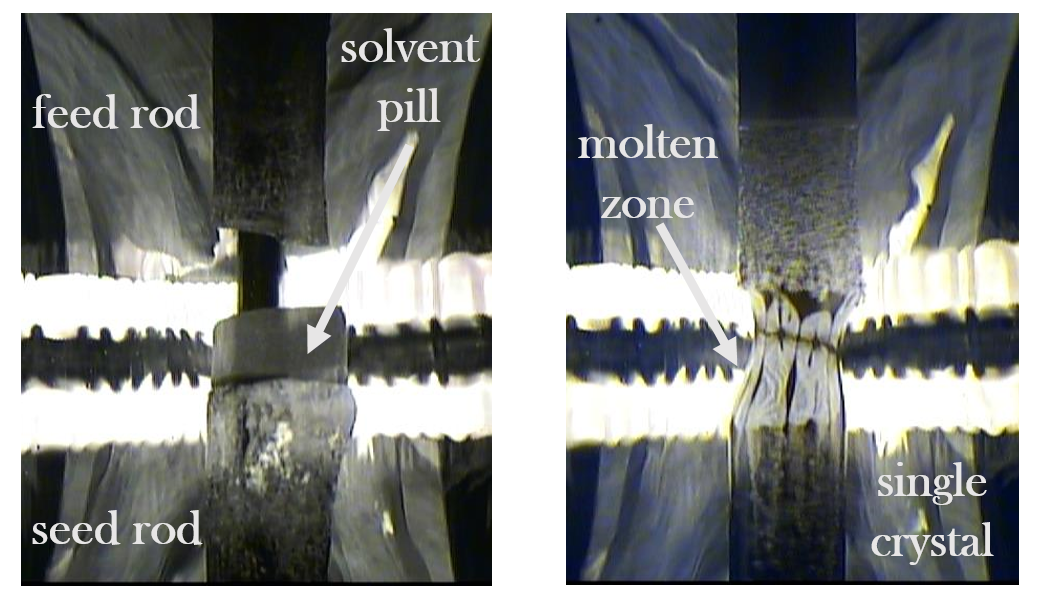}
    \caption{Images from inside the mirror furnace. {\it Left:} Initial configuration showing the three main components, the feed and seed rods, the solvent pill, and the bright light radiating from the halogen lamps. {\it Right:} Snapshot taken during a successful growth picturing the molten zone and the solidified single crystal material.}
    \label{fig:crystal growth}
\end{figure}

The sample used throughout this study is a cylindrically shaped (17.9~mm height, 4.75~mm diameter and $\sim$5~g mass) non-annealed LSCO sample with nominal Sr doping x = 0.08. The crystal has been grown using the Travelling Solvent Floating Zone technique (TSFZ) \cite{tanaka1989single} at Department of Energy Conversion and Storage, Technical University of Denmark. In the field of cuprates superconductors the TSFZ method is preferred to other procedures, such as the flux method or the Czochralski method, for a number of reasons. Firstly, the crystallisation takes place inside a mirror furnace thus preventing contamination which might arise from the chemical reaction between crucibles and the melt \cite{bhat2014introduction}. As shown in Figure \ref{fig:crystal growth}, the polycrystalline feed rod together with the solvent pill (which acts as a promoter of the molten zone) are vertically melted by very powerful halogen lamps. After a stable molten zone has been achieved, the lamps are being translated upwards (or the feed rod downwards) allowing the material to solidify and, if the right thermodynamic conditions are achieved, crystallise into a single crystal phase. At the same time, TSFZ is the only of the three above-mentioned techniques which permits the growth of centimeter-size single crystals of compounds that melt incongruently \cite{le2011phase}. One of the disadvantages that might arise from using this method to synthesise LSCO samples in particular is the lack of compositional homogeneity of the melt throughout the growth process. This combined with the fact that the Sr atoms are fixed after crystallisation, unlike compounds with mobile dopants such as oxygen doped La$_2$CuO$_{4+y}$, results in heterogeneous samples \cite{oka1994crystal}. Because it is impossible to monitor or control the dopant concentration during the growth, it is important to characterise the entire bulk of the sample before proceeding to draw conclusions on the fundamental properties of this family of superconductors. 

\begin{figure}
    \centering
    \includegraphics[width=0.7\columnwidth]{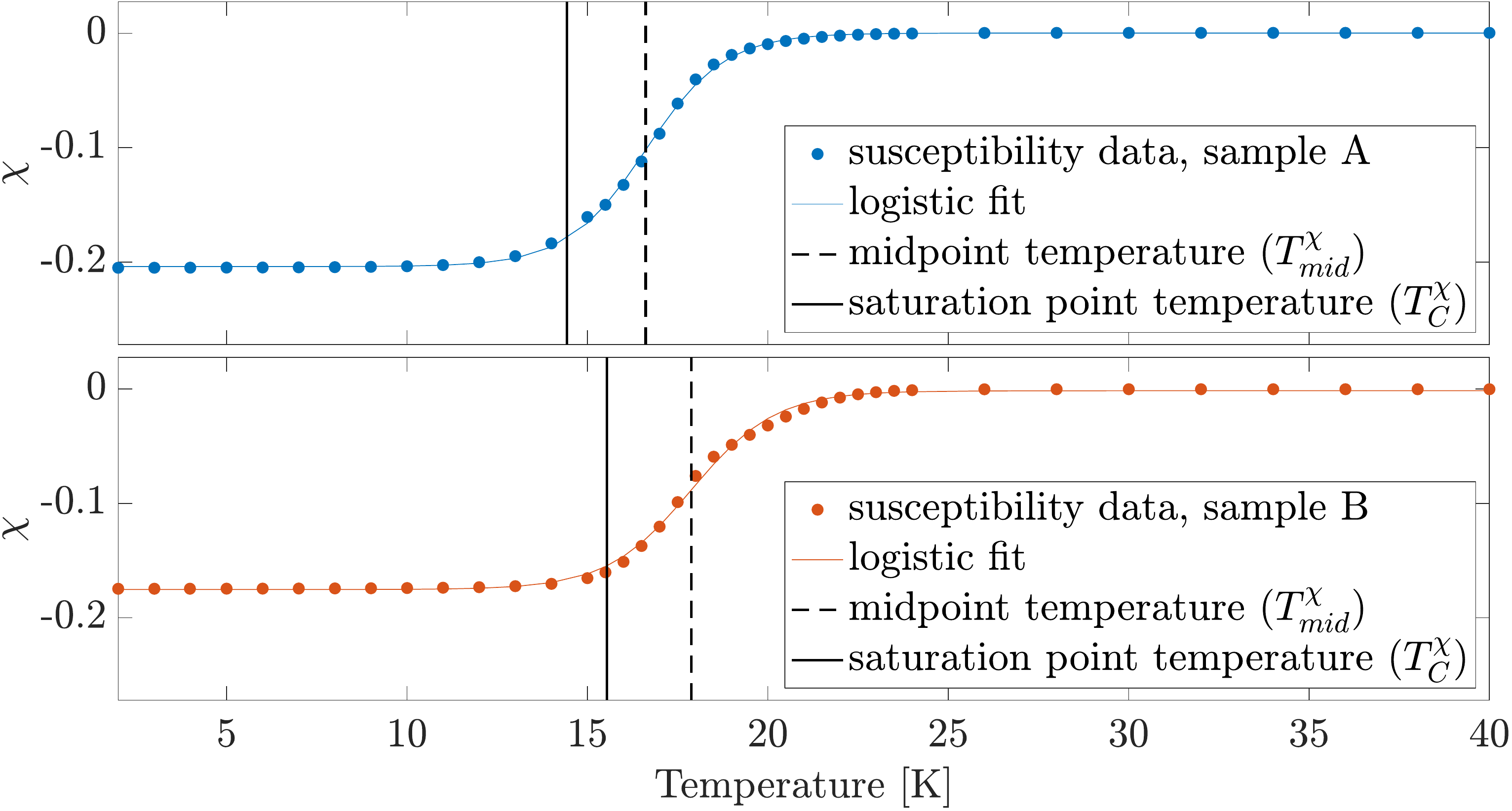}
    \caption{Magnetic susceptibility data of the two the cubes cut from each of the ends (A and B) of the single crystal sample. The deviation from $\chi = -1$ at low temperatures is due to magnetic flux trapping, which occurs when cooling the sample through the phase transition in an applied magnetic field H $= 800$ A/m ($10$ Oe). See Figure~\ref{fig:fitex} and eq.~\ref{eq:logistic} for fitting procedure.}
    \label{fig:chi}
\end{figure}

There are numerous ways to determine the doping concentration of a LSCO sample, some of the most widely used ones being measurements of magnetic susceptibility \cite{radaelli1994structural}, electrical resistivity \cite{ando2004electronic}, heat capacity \cite{momono1994low} and of the structural phase transition between the high temperature tetragonal phase and the low temperature orthorhombic one by means of elastic neutron scattering \cite{wakimoto2004neutron}. In order to test the doping homogeneity of our crystal, we have performed temperature dependent magnetic susceptibility measurements using a Quantum Design MPMS SQUID magnetometer. The set-up was used in DC magnetic measurements mode where the magnetic moment of the sample is determined from the local changes, induced by the movement of the sample, in the applied magnetic flux. Taking into account the sample volume, the magnitude of the applied magnetic field as well as demagnetising effects, the measured magnetic moment can be converted into units of susceptibility ($\chi$) \cite{McElfresh1994}. From the susceptibility as a function of temperature so obtained, it is possible to determine the critical superconducting temperature (T$_\text{C}$) at which the material abruptly transitions from a small positive susceptibility, corresponding to the antiferromagnetic state, to a diamagnetic phase characterised by negative susceptibility. T$_\text{C}$ can be defined in various ways, the onset of the transition and the midpoint being the most popular ones. Throughout this paper we define the magnetic susceptibility critical temperature ($T_\text{C}^{\chi}$) as the diamagnetic saturation temperature obtained by fitting the data to a logistic function and following the slope at the inflection point down to the lower asymptote as it can be seen in Figure \ref{fig:chi}. This corresponds to the temperature at which a fully coherent superconducting state is established in the sample. 

Previous experimental studies have revealed a strong critical temperature anisotropy in LSCO superconductors. Measurements performed with applied magnetic field parallel to the a-b plane show a higher critical temperature as well as an unexpected paramagnetic Meissner effect, recorded in the field cooled process under low applied magnetic fields \cite{felner2013anisotropy}. To avoid this complication and in order to be able to directly compare our results with the ones from the literature \cite{adachi2009magnetic} we chose to measure the magnetic susceptibility in the c-direction by using an applied magnetic field, H $= 800$ A/m ($10$ Oe), parallel to the c-axis (H$\|$c). For this the sample was aligned by hard X-ray diffraction, and small cubic pieces ($\sim$ 1 mm length) were cut from the two ends of the large single crystal, such that the c-axis was parallel to one of facets. The $\chi(T)$ measurements were performed using the field cooling (FC) protocol, where the sample is cooled from a temperature above T$_\text{C}$ in a small static magnetic field (H = $800$ A/m ), while continuously recording data.  Around the predicted critical temperature, from 24~K to 15~K, measurements have been taken every 0.5~K, while outside this range, larger temperature steps have been employed, 2~K steps from 40~K to 24~K and 1~K steps from 15~K down to 2~K. The measurement has revealed a mismatch in transition temperature between the two samples originating from the two ends of the main single crystal. One piece had a transition temperature, $T_\text{C}^{\chi}$, of $14.43 \pm 0.09$~K and the other a $T_\text{C}^{\chi}$ of $15.54 \pm 0.18$~K (See Figure~\ref{fig:chi}). This mismatch in critical temperature motivated the search for a non-destructive method that would be able to reveal doping variations in the entire bulk of the sample.

\section{Polarised neutron imaging}
\begin{figure}
    \centering
    \includegraphics[width=0.6\columnwidth]{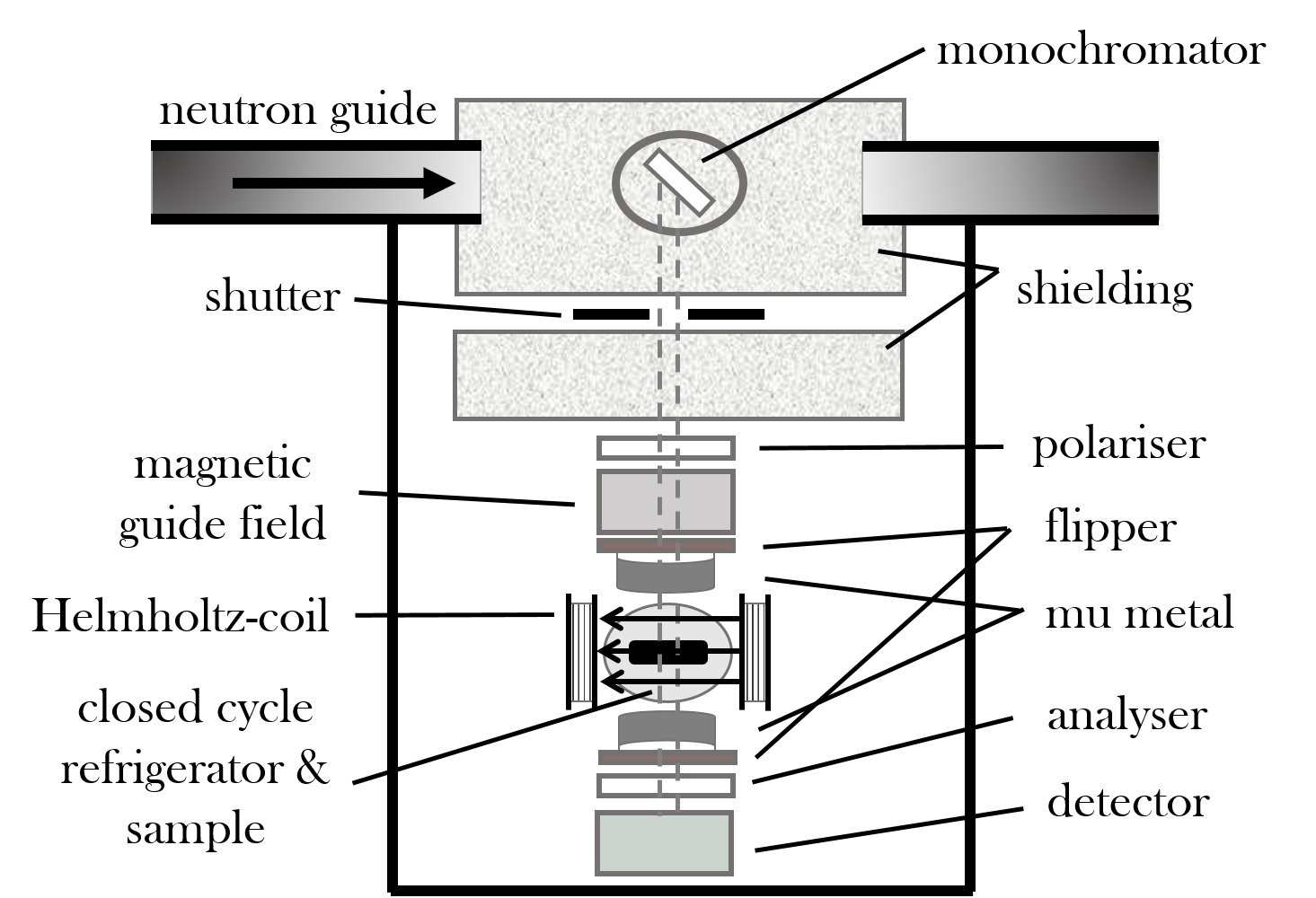}   
    \caption{Sketch of the polarised neutron set-up used for the present measurements. Adapted from \cite{treimer2014radiography}.}
    \label{fig:instrument}
\end{figure}

In order to probe the doping distribution along the entire length of the crystal in a non-destructive manner, we have performed polarised neutron imaging experiments at the PONTO-III (V21) instrument at the BER II nuclear reactor in Helmholtz-Zentrum Berlin. The instrumental set-up is shown in Figure~\ref{fig:instrument}. The instrument makes use of a cold neutron beam which, for this experiment, was monochromatized to 3.4 \AA{}. 

The control of the initial spin state of the neutrons as well as the analysis of their final spin state was done with the help of two super-mirror polarisers placed before and after the sample. A guide field was used to keep the neutron spin vertically aligned from the polariser up to the sample region and mu-metal shielding was used just before the sample as well as between the sample and the analyser. A single spin-flipper was used to select between spin-up and spin-down measurements.

In a polarised neutron imaging experiment there are two main methods of determining the superconducting critical temperature: by probing the Meissner transition or by imaging the trapped magnetic flux inside the superconductor as a function of temperature \cite{treimer2014radiography, kardjilov2008three}. In this study we employed the latter method. 
This requires cooling the superconductor through its critical temperature inside an applied magnetic field and performing the measurements while heating up the sample without applied field. This way, one is able to trap magnetic flux inside the superconductor and probe its release as the sample returns to the normal, non-superconducting state. 

The single crystal used in this study was aligned in the $(b-c)$ plane and mounted on an aluminium sample holder inside a closed cycle refrigerator which was able to reach temperatures as low as 4.5~K. During the cooling procedure, a pair of Helmholtz coils were used to apply a horizontal magnetic field parallel to the c-axis of the sample and perpendicular to the neutron beam direction. The sample was slowly cooled from room temperature down to 100 K with a rate of 2~K/min followed by a higher cooling rate of 5~K/min down to 40~K. An external magnetic field was applied and the sample was further cooled below its superconducting critical temperature down to the base temperature of the closed cycle refrigerator, thus trapping the magnetic flux inside the sample. 
The external magnetic field was then switched off and radiographs were recorded, by a  scintillator - CCD camera detection ensemble (camera: Andor DV $434$-BV, effective pixel size: $77$~$\upmu$m, $^6$LiFZnS scintillator thickness: $400$~$\upmu$m \cite{kardjilov2011highly}), at various temperatures during the heating process. 
The horizontal divergence of the neutron beam was $0.34^{\circ}$ and the vertical divergence was $1.28^{\circ}$.

Each image registers the polarisation of the neutron spin, in modulo 360$^{\circ}$
, analysed by the second super-mirror polariser placed between the sample and the scintillator. As the neutron passes through the magnetic field trapped inside the superconducting material, the spin will undergo a so-called Larmor precession. The precession angle ($\phi$) is dependent on the strength of the applied magnetic field and the time spent by the neutron inside it, which  in turn is dependent on the neutron velocity (and thereby on its wavelength) as well as the length of the path through the magnetic field \cite{treimer2014radiography}:

\begin{align}
    \phi = \frac{\gamma_{L} m}{h}BL\lambda .
\end{align}
Here, 
$\gamma_{L} = {-1.83247179(43)} \times 10^8$
~rad~s$^{-1}$~T$^{-1}$ 
is the neutron's gyromagnetic ratio, 
$h = 6.62607015 \times 10^{-34}$~
J\,s is Plank's constant, 
$m=1.674927471(21) \times 10^{-27}$~kg is the mass of a neutron and $B, L$ and $\lambda$ are the strength of the applied magnetic field, the length of the neutron trajectory inside the field and the neutron wavelength respectively. 

The magnitude of the applied magnetic field was of the order of a few milli-tesla, and was tuned to provide the most contrast while keeping the maximum precession angle below 180$^{\circ}$ in order to avoid ambiguity, as the measured polarisation is the cosine of the (ensemble average of the) neutron spin precession angle:
\begin{align}
    P=\langle\cos(\phi)\rangle
\end{align}
In our experiment the maximum measured precession angle was $\sim105^{\circ}$.

Radiographs were recorded in two spin configurations: spin up and spin down, with the spin of the neutron in both being perpendicular to the beam direction. Spin down was selected by using an additional magnetic field to add a neutron spin precession angle of 180$^{\circ}$ before the analyser. 
For normalisation purposes, dark frame images (DF) without neutrons and in
zero applied magnetic field were recorded at room temperature (which means no trapped field in the sample) so that noise in the detector systems could be corrected for. Open beam images (OB) were recorded as well at room temperature, in the absence of both the sample and the applied magnetic field, so that imperfections in the polarisation setup could be corrected for. 

The dark frame radiographs were subtracted from all the corresponding images, including the open beam ones. 
The measuring time was 10~minutes per image (composed of 4 exposures of 150~s combined after DF subtraction).
The open beam polarisation was calculated as:

\begin{align}
    P_{OB} = \frac{\uparrow_{OB} - \downarrow_{OB}}{\uparrow_{OB} + \downarrow_{OB}}
\end{align}

where $\uparrow$ and $\downarrow$ indicate the radiographs taken with spin up and down configuration, respectively. The uncorrected polarisation signal with sample, $P_{UC}$, was measured and calculated in the same way,
whereafter the open beam corrected polarisation, $P$, was calculated as:
\begin{align}
    P = \frac{P_{UC}}{ P_{OB}}.
\end{align}

%
%
%

\begin{figure}
    \centering
    \includegraphics[width=0.7\columnwidth]{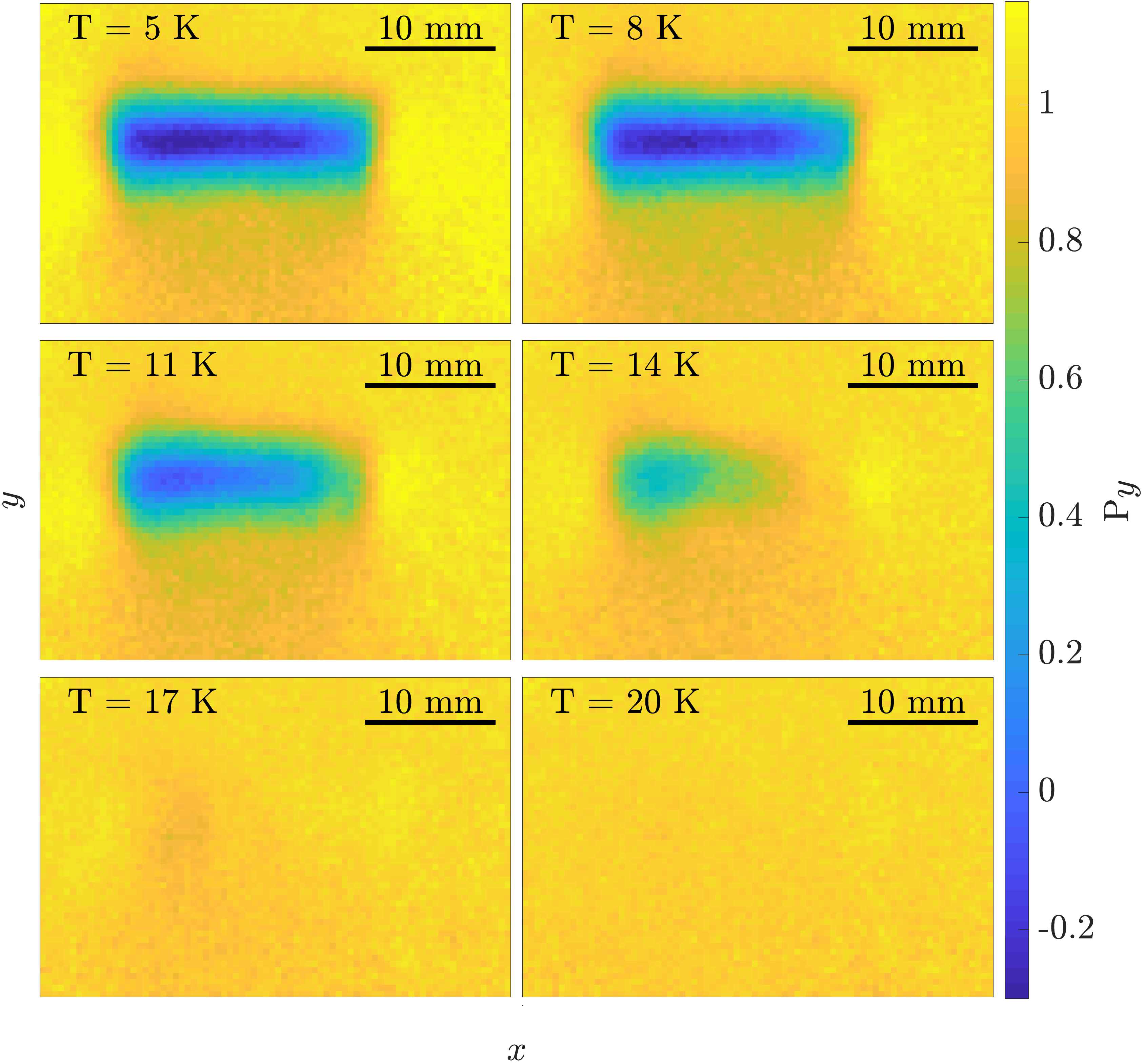}
    \caption{Polarisation transmission images as a function of temperature in the range 5~K to 20~K. The sample position (blue spot) can be seen in 5~K plot where neutrons passing through the trapped magnetic field in the sample undergo precession thereby moving the spin directions away from their initial polarisation along $y$. Neutrons were travelling along $z$ and probing the trapped magnetic field, which was along $x$. As the temperature is raised the trapped field dissipates and no precession signal is observed. At 20~K the trapped field is completely gone. Notice how different ends of the sample lose their trapped field at different temperatures.}
    \label{fig:PvsT}
\end{figure}

\section{Results and discussion}
\subsection{The polarised imaging data}
Figure~\ref{fig:PvsT} shows an image of the polarisation of the beam transmitted through the sample, collected at six temperatures between 5~K and 20~K. In total, we collected data in temperature steps of 1~K between 5~K and 34~K. The direction of the field was along $x$, the neutrons were initially polarised along $y$, and the polarisation analysis was also performed along $y$. To improve counting statistics, the data was rebinned spatially by a factor of 6 along $x$ and $y$ to an effective pixel size of $0.462\times 0.462$~mm$^2$. At the lowest temperature it can be seen that the entire LSCO sample causes a large deviation of the neutron beam polarisation vector with respect to the initial imposed direction (blue colour). In the sample centre, the polarisation even reaches negative values, indicating a coherent rotation of the neutron polarisation inside the sample. This we interpret as the result of a trapped magnetic field. 
When the temperature is increased, the release of the trapped magnetic field is found to take place at different temperatures for different parts of the sample, as is evident from Figure~\ref{fig:PvsT}. In particular, at 14~K only the left part of the sample depolarises the beam (i.e.\ is still in the superconducting state). 

\subsection{Determining the distribution of $T_\text{C}$}
\begin{figure}
    \centering
    \includegraphics[width=0.5\columnwidth]{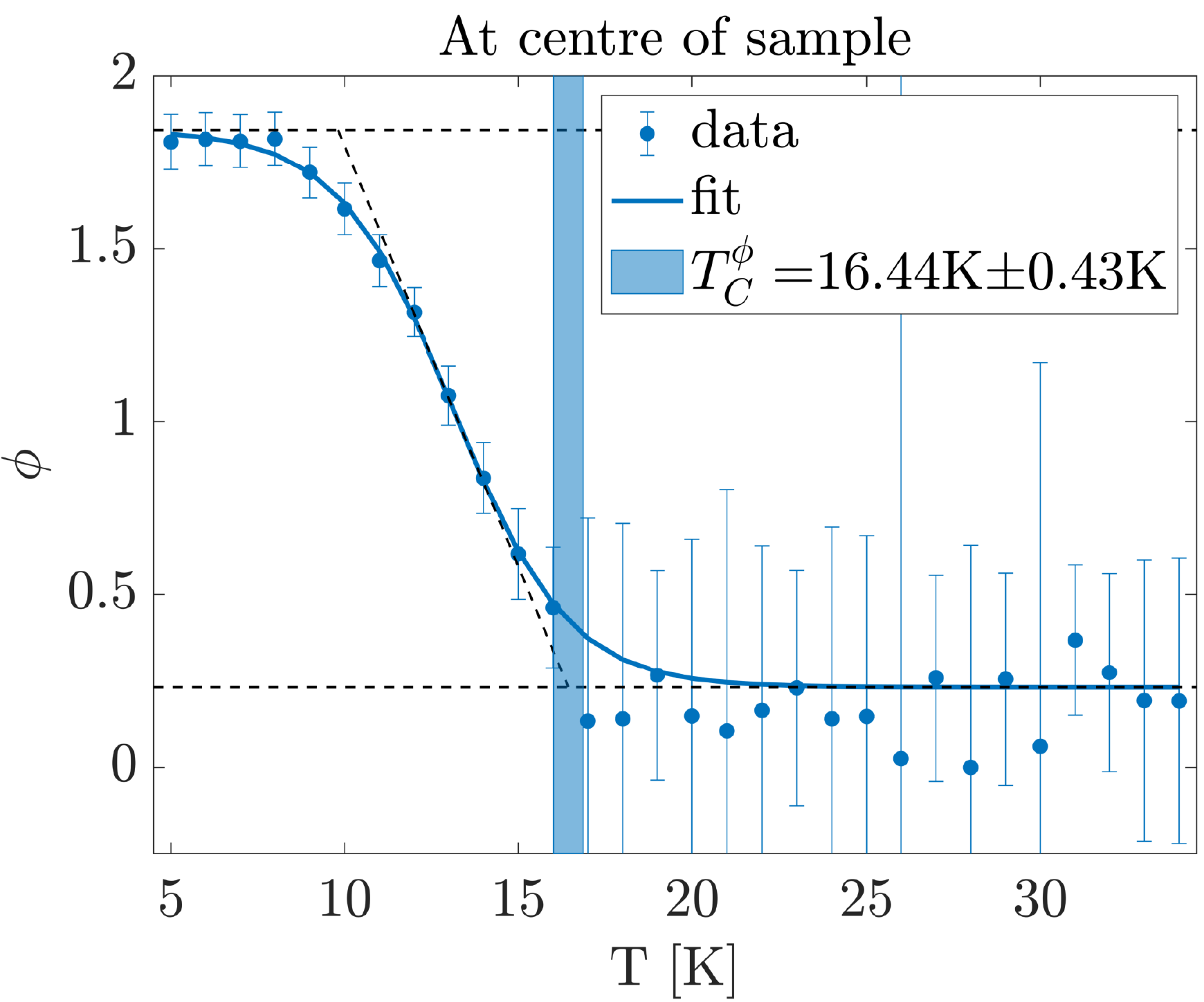}
    \caption{Example showing the fitting procedure for the central pixel of the sample using equation~\ref{eq:logistic} to find the transition temperature, $T_\text{C}$, indicated by the shaded area.}
    \label{fig:fitex}
\end{figure}
In order to quantitatively assess the superconducting transition temperature, $T_\text{C}$, for each pixel within the sample region, we first convert the polarization data to spin precession angle through $\phi=\arccos{(P)}$, for each temperature, $T$.  Values for pixel polarisations that due to noise were (slightly) larger than 1 were set to 1. Next, a logistic curve was fitted to the precession angle data, as the logistic function describes the sigmoidal nature of the phase transition between superconducting and non-superconducting states, such that values for the critical transition temperature can be extracted:
\begin{align} \label{eq:logistic}
    \phi=\phi_{\textrm{min}}+\frac{\phi_{\textrm{max}}-\phi_{\textrm{min}}}{1+\textrm{e}^{-n_H\left(T-T_{\textrm{mid}}  \right)}}.
\end{align}
$\phi_{\textrm{min}}$ and $\phi_{\textrm{max}}$ are the low and high asymptotes, respectively, $T_{\textrm{mid}}$ is the temperature at the inflection point of the slope, and $n_H$ is Hill's coefficient, giving the steepness of the curve. 

A fitting example for the centre-most pixel of the sample is shown in Figure~\ref{fig:fitex}.
We define the critical temperature, $T_\text{C}^{\phi}$, as the temperature where the trapped field is fully released. Within the logistics model, we approximate this by following the slope at the inflection point down to the lower asymptote as indicated by the dotted lines in the figure. The shaded area indicates the uncertainty interval of the estimated $T_\text{C}^{\phi}$, where the error estimate was obtained from error propagation of the errors from the least square fit of the data, which was weighed using the error from the counting statistics of the measurement. Note that the low asymptote is non-zero, since a background magnetic field from the experimental set-up was present for the experiments at the given temperatures.

\begin{figure}
    \centering
    \includegraphics[width=1\columnwidth]{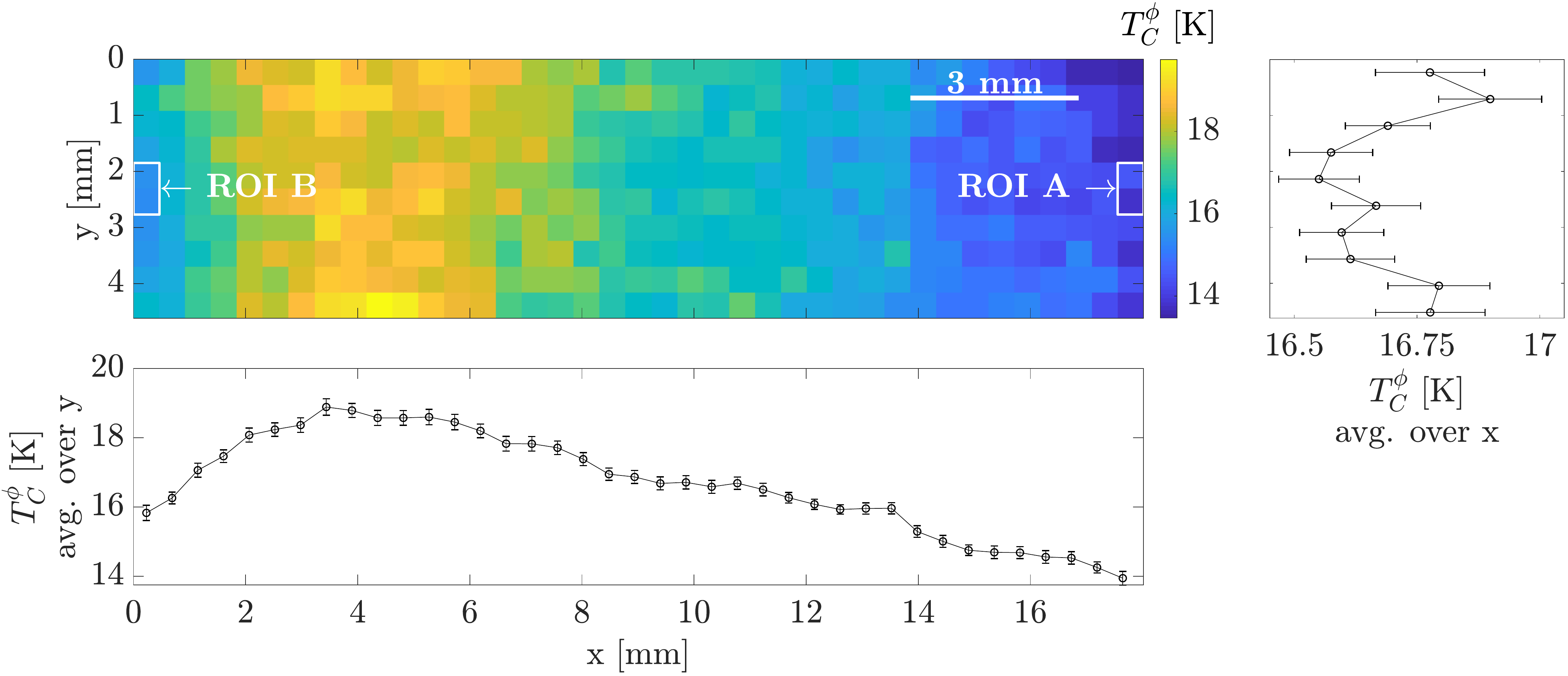}
    \caption{{\it Top left}: The transition temperature mapped out across the sample. Found using the method presented in Figure~\ref{fig:fitex}. The regions of interest (ROIs) indicated are used for comparing to susceptibility measurements on sample pieces that were adjacent to the ROIs (See Figure~\ref{fig:phivschi}). {\it Bottom:} Transition temperature along the length of the sample, with data averaged over the width of the sample. {\it Top right:} Transition temperature along the width of the sample, with data averaged over the length of the sample. Note that, due to the vertical divergence, quantitative conclusions on the radial variation are limited.}
    \label{fig:2Dand1D} 
\end{figure}

The same fitting procedure was used to find $T_\text{C}^{\phi}$ for every pixel covered by the sample. The result is shown in Figure~\ref{fig:2Dand1D}. A clear variation of $T_\text{C}^{\phi}$ is observed along the length of the sample, with a maximum $T_\text{C}^{\phi}$ of about 19~K at approximately 4~mm from one sample end. $T_\text{C}^{\phi}$ then drops to about 16~K at the end closest to the maximum and to 14~K at the opposite end. This corresponds to a doping range $x \simeq 0.075 - 0.082$ \cite{radaelli1994structural,kofu2009hidden}. The doping range is in agreement with the nominal doping of the crystal, its average value being exactly the intended doping value ($x=0.080$) of the initial powder mixture. The variation across the length of the crystal can be explained as a slight accumulation of strontium atoms during one stage of the growth which in turn caused a deficiency of dopant at other stages. Furthermore, our results suggest that there is a radial variation of Sr distribution with more Sr at the sample surface, however, due to limited vertical resolution this result warrants further investigation. The required resolution is available at other neutron imaging instruments such as RADEN at J-PARC MLF \cite{Shinohara2020}.    

In order to rule out a possible temperature gradient, the polarised imaging data was analysed to confirm that the temperature had stabilised before exposures and that no fluctuations were seen afterwards. Previous studies, with the same cryogenic setup, mention temperature variations of maximum 0.01~K \cite{kardjilov2008three}.

Using the polarised neutron imaging method we were able to identify the precise pattern of the critical temperature gradient, and implicitly the doping distribution, along the entire 17.9 mm length of the sample, in the growth direction. Considering that, within the TSFZ method, the molten zone is continually mixed and the material crystallises layer by layer, we do not expect a significant doping gradient in the radial direction. This can however be investigated using the tomography capabilities of a polarised imaging set-up. 
One of the advantages of using highly underdoped LSCO samples for this type of investigation is the fact that in this region of the phase diagram T$_\text{C}$ varies steeply with doping, meaning that a small variation in Sr concentration corresponds to a significant critical temperature difference. However, using a high accuracy temperature control one would be able to determine the doping variation in sample across the entire superconducting dome, except the range $x=0.14 - 0.18$, where $T_{\rm C}$ is essentially constant.


\subsection{Validation of the polarised imaging method}

\begin{figure}
    \centering
    \includegraphics[width=0.7\columnwidth]{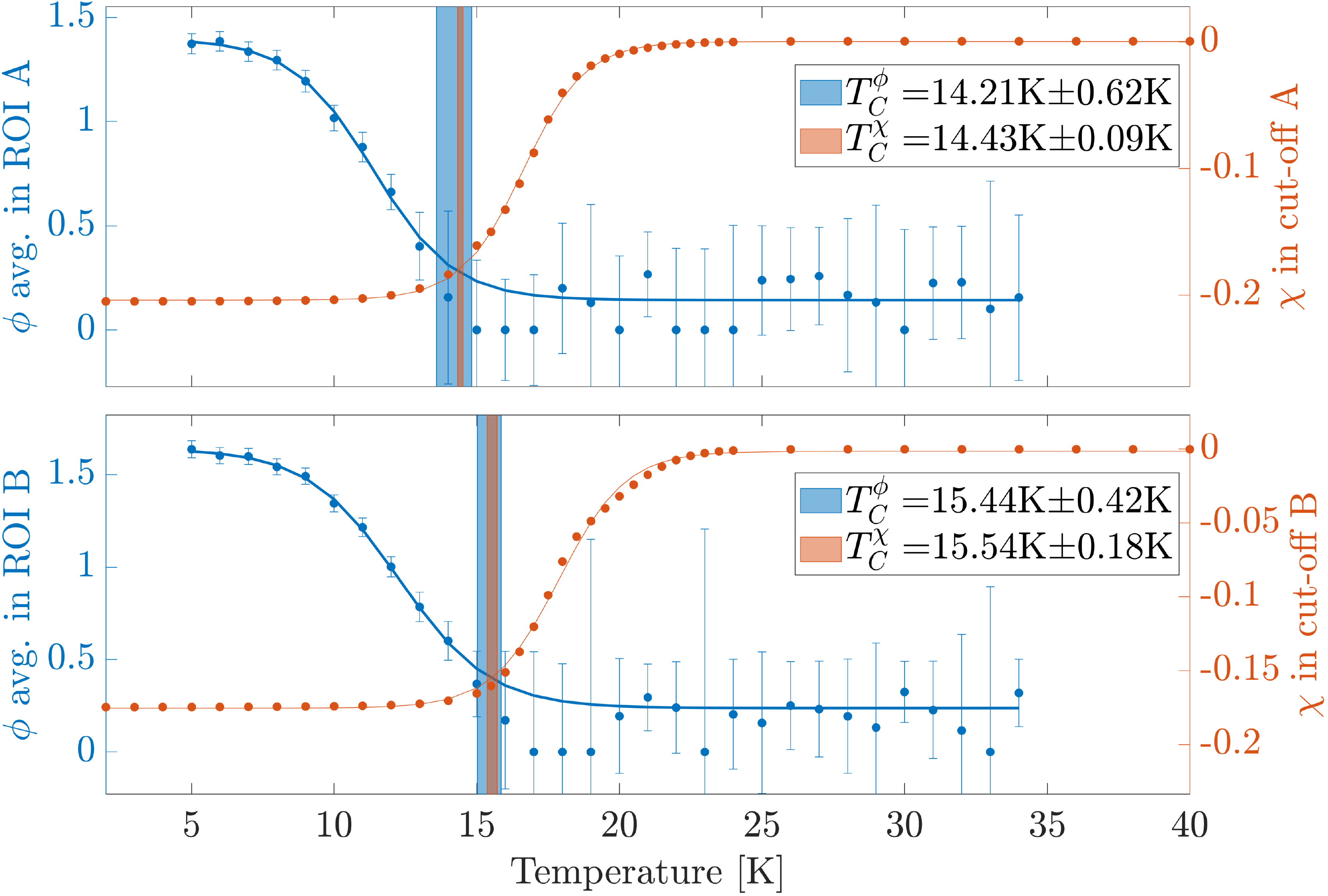}
    \caption{{\it Blue}: Fit of precession angle, $\phi$, obtained by measuring polarised neutron spin precession in a trapped magnetic field as function of temperature. {\it Red}: Fit of susceptibility, $\chi$, measured on cut off sample pieces that were adjacent to the two ROIs as indicated in Figure~\ref{fig:2Dand1D}, ROI~A ({\it Top}) and ROI~B ({\it Bottom}).}
    \label{fig:phivschi}
\end{figure}

We have also benchmarked our imaging results against magnetic susceptibility ($\chi(T)$) data. In Figure~\ref{fig:phivschi}, we show, in comparison, the susceptibility data and the measurement of polarised neutron spin precession in trapped magnetic field for the two ends of the sample (shown as the two panels in the figure). As seen, the results from the two measurements are in agreement, if we define $T^{\chi}_\text{C}$, from the susceptibility data, to be the temperature where the negative susceptibility is saturated. Within errorbars, we find the same difference between the two sample ends for the two measurement methods. Due to the fact that the two methods are mapping different phenomena (one follows the diamagnetic response of the superconductor while the other images the expulsion of the trapped magnetic field as a function of temperature) a precise definition of T$_\text{C}
$, to be applied to both measurements, is very hard, if not impossible, to formulate. However, the great agreement we are able to obtain by using the same 
fitting routine on the two data sets indicates that a fully developed 
superconducting state is a prerequisite for the formation of a vortex 
lattice and it proves that polarised neutron imaging is a reliable 
method of determining the spatial variation of the superconducting 
critical temperature and thus the doping level of superconducting 
crystals.

Where the susceptibility method is more accurate (uncertainty around 0.1~K) than the polarised imaging method (uncertainty around 0.5~K), the latter method yields a complete picture of the sample and is furthermore non-destructive. For this reason, we can imagine polarised imaging to be an interesting alternative to susceptibility measurements for investigating the spatial variations in transition temperatures, and doping implicitly, in samples of centimetre sizes. Furthermore, by upgrading the instrumental components and extending the exposure time (and/or neutron flux) the precision of the measured transition temperature can be improved as it is currently limited by counting statistics. The spatial resolution can be improved as well down to 100~$\mu m$ \cite{strobl2019polarization}.

\section{Conclusion}
Homogeneity of a sample is important when investigating the superconducting properties, especially when performing bulk measurements. We have shown that through spin-polarised neutron imaging it is possible to directly map out the spatial distribution of the critical temperature thereby providing a characterisation method capable of eliminating uncertainties and sources of error related to -- in this case -- variations of the distribution of Sr doping in a La$_{2-x}$Sr$_x$CuO$_4$ high temperature superconductor. The results have been substantiated by magnetic susceptibility measurements showing the same trend for the $T_\text{C}$ variation between the two ends of the sample.

Furthermore, as we are simply measuring a scalar property, our presented method for mapping the $T_\text{C}$ distribution, can easily be expanded to three dimensional investigations using standard filtered backprojection methods for tomographic reconstructions.

\section*{Acknowledgements}
We would like to thank the Helmholtz Zentrum Berlin for granting us beam time to perform the polarised imaging studies. The neutron scattering experiments were supported by the Danish Agency for Research and Innovation through DANSCATT. A.-E. \c{T}u\c{t}ueanu was supported through the Institute Max von Laue Paul Langevin (ILL) Ph.D. program. Y. Sassa was fully supported by a VR neutron project grant (BIFROST, Dnr. 2016-06955) as well as a VR starting grant (Dnr. 2017-05078). 
The authors would like to thank Dr. Pierre Courtois and Philipp Gausmann, from the ILL, for their help in cutting and aligning the crystals in preparation for the magnetic susceptibility measurements.

\section*{Author Contributions}
The two authors, A.-E. \c{T}u\c{t}ueanu and M. Sales, contributed equally to this work.

\clearpage
\bibliography{references.bib}

\end{document}